\documentstyle[pra,aps]{revtex}
\begin{document}
\title{Optimum Unambiguous Discrimination Between Linearly Independent Symmetric States}
\author{Anthony Chefles\thanks{e-mail: tony@phys.strath.ac.uk} and Stephen M. Barnett}
\address{Department of Physics and Applied Physics, \\ University of Strathclyde,
       Glasgow G4 0NG, Scotland}
\input epsf
\epsfverbosetrue

\def\id{\hat{\leavevmode\hbox{\small1\kern-3.2pt\normalsize1}}}%

\maketitle

\begin{abstract}
The quantum formalism permits one to discriminate sometimes between any set of linearly-independent pure states with certainty.  We obtain the maximum probability with which a set of equally-likely, symmetric, linearly-independent states can be discriminated.  The form of this bound is examined for symmetric coherent states of a harmonic oscillator or field mode.
\end{abstract}

\section{Introduction}
\renewcommand{\theequation}{1.\arabic{equation}}
\setcounter{equation}{0}

It is possible to manipulate the state of a quantum system in far more interesting ways than can be achieved by carrying out unitary operations and von Neumann measurements on the system of interest alone.  Consideration of the effects of interactions with other systems has led to the development of the quantum operations formalism\cite{Kraus}, which allows any completely positive, trace-preserving map to represent, in principle, a realisable transformation of the density operator ${\hat \rho}$.  One particularly interesting type of operation is a probablistic operation.  This is an operation which, with some probability  less than 1, will transform the state of the system in a manner which cannot be brought about by any deterministic process.  Although such operations generally have a non-zero failure probability, one generally knows whether or not the desired transformation has taken place.  

An important class of probablistic operations are those which allow one to discriminate unambiguously between non-orthogonal states, that is, with zero probability of error.  When carried out on a quantum system prepared in one of the non-orthogonal states $|{\psi}_{j}{\rangle}$, such an operation will, with some probability, transform the state into a corresponding member of an orthonormal set $|{\phi}_{j}{\rangle}$.  The latter states can be discriminated without error using a simple von Neumann measurement.  Although such an operation cannot have unit probability of success,  we can always tell whether or not the desired transformation has taken place.  When the attempt fails, we obtain an inconclusive result.

The subject of unambiguous state discrimination was pioneered a decade ago by Ivanovic\cite{Ivanovic}, Dieks\cite{Dieks} and Peres\cite{Peres1}, and has recently undergone interesting further developments.  While earlier work concentrated only on the problem of discriminating between two states, the problem of discriminating between multiple states has since been addressed.  In particular, one of us\cite{Me1} has shown that the necessary and sufficient condition for a set of states $|{\psi}_{j}{\rangle}$ to be amenable to unambiguous state discrimination is that they must be linearly-independent.  More recently, the problem of discriminating between three states has been examined in detail by Peres and Terno\cite{Peres2}.  We have shown that unambiguous discrimination is intimately related to other well-known types of probablistic operation, such as entanglement concentration\cite{Me1,Me2} and exact cloning\cite{Me3}.  The latter connection has also been examined by Guan and Duo\cite{Guanduo}.  Unambiguous discrimination between two non orthogonal states has been demonstrated in the laboratory by Huttner {\em et al}\cite{Huttner}.  In this experiment, weak pulses of light were prepared in non-orthogonal polarisation states, a fraction of which were converted into orthogonal ones by a loss mechanism.

It is clearly of interest to find the optimum strategy for discriminating unambiguously between a set of known states, that is, to determine the maximum probability of obtaining a definite result.  As with problems in conventional quantum detection theory\cite{Helstrom}, where the aim is to find the {\em absolute} maximum of the discrimination probability or mutual information for a given source, few analytic solutions for optimum strategies and their respective figures of merit are known.  The complete solution for unambiguous discrimination between two states with arbitrary {\em a priori} probabilities has been found by Jaeger and Shimony\cite{JShimony}.  Peres and Terno\cite{Peres2} have explored the geometry of the optimisation problem for three states, and obtained useful insight into the general $N$ state case.  As yet, however, no analytical solutions have been found for more than two states.  Such a solution is given in this paper.  We determine the maximum probability with which $N$ symmetric states can be unambiguously discriminated, assuming they have equal prior probabilities.   We then apply our result to examine the maximum probability of discriminating between $N$ symmetric coherent states $|{\alpha}_{j}{\rangle}$. 

\section{Unambiguous State Discrimination}
\renewcommand{\theequation}{2.\arabic{equation}}
\setcounter{equation}{0}
Consider a quantum system prepared in one of $N$ pure quantum states $|{\psi}_{j}{\rangle}$, where $j=0,...,N-1$.  These states span an $N$-dimensional Hilbert space  ${\cal H}$.  If the states are non-orthogonal, no quantum operation can deterministically discriminate between them.  It is, however, possible to devise a strategy which, with some probability, will reveal the state with zero error probability.  To see how this may be done, it is convenient to employ the Kraus representation of quantum operations\cite{Kraus}.  Each of the possible, distinguishable outcomes of an operation, which are labelled by the index ${\mu}$, is associated with a linear transformation operator ${\hat A}_{\mu}$.  These form a resolution of the identity

\begin{equation} 
\sum_{\mu}{\hat A}^{\dagger}_{\mu}{\hat A}_{\mu}={\id}.
\end{equation} 
If the system is prepared with the initial density operator ${\hat
\rho}$, the probability $P_{\mu}$ of the $\mu$th outcome is
${\mathrm Tr{\hat \rho}}{\hat A}^{\dagger}_{\mu}{\hat A}_{\mu}$.
The final density operator corresponding to this result is ${\hat A}_{\mu}{\hat \rho}{\hat
A}^{\dagger}_{\mu}/P_{\mu}$.
The state discrimination operation will have $N+1$ distinct outcomes, corresponding to detection of each of the states, and an additional answer which gives no information about the state.  This is the inconclusive result. The operator which corresponds to the detection of the state $|{\psi}_{j}{\rangle}$ is ${\hat A}_{j}$, where $j=0,...,N-1$, and we let ${\hat A}_{F}$ be the operator which leads to a failure of the discrimination attempt.  Clearly, we have

\begin{equation} 
{\hat A}^{\dagger}_{F}{\hat A}_{F}+\sum_{j}{\hat A}^{\dagger}_{j}{\hat A}_{j}={\id}.
\end{equation}  
The zero-errors condition takes the form
\begin{equation}
{\langle}{\psi}_{j}|{\hat A}^{\dagger}_{j'}{\hat A}_{j'}|{\psi}_{j}{\rangle}=P_{j}{\delta}_{jj'},
\end{equation}
where $P_{j}$ is the conditional probability, given that the system was prepared in the state $|{\psi}_{j}{\rangle}$, that this state will be identified.  This zero-errors condition can only be met if the $|{\psi}_{j}{\rangle}$ are linearly-independent, and we find that the ${\hat A}_{j}$ have the form\cite{Me1}:
\begin{equation}
{\hat A}_{j}=\frac{P^{1/2}_{j}}{{\langle}{\psi}^{\perp}_{j}|{\psi}_{j}{\rangle}}|{\phi}_{j}{\rangle}{\langle}{\psi}^{\perp}_{j}|,
\end{equation}
where the $|{\phi}_{j}{\rangle}$ form an orthonormal basis for ${\cal H}$.  Here we have also introduced the {\em reciprocal states} $|{\psi}^{\perp}_{j}{\rangle}$.  The reciprocal state $|{\psi}^{\perp}_{j}{\rangle}$ is defined as that which lies in ${\cal H}$ and is orthogonal to all $|{\psi}_{j'}{\rangle}$ for $j{\neq}j'$.  The set of  states $|{\psi}^{\perp}_{j}{\rangle}$ is simply the $N$-dimensional complex generalisation of the set of (normalised) reciprocal vectors in crystallography \cite{Guinier} with respect to the unit cell basis vectors, these being the $|{\psi}_{j}{\rangle}$.  A complete set of reciprocal states exists if, and only if, the $|{\psi}_{j}{\rangle}$ are linearly-independent.  The reciprocal states are also necessarily linearly-independent, as is shown in \cite{Me1}.

Given that the states $|{\psi}_{j}{\rangle}$ have {\em a priori} probabilities ${\eta}_{j}$, the total probability of correctly identifying the state is
\begin{equation}
P_{D}=\sum_{j}{\eta}_{j}P_{j}=\sum_{j}{\eta}_{j}{\langle}{\psi}_{j}|{\hat A}^{\dagger}_{j}{\hat A}_{j}|{\psi}_{j}{\rangle}.
\end{equation}
It convenient to proceed using the language of Positive Operator-Valued Measures (POVMs)\cite{Kraus}.  The measurement can be expressed as an $N+1$ element POVM operation by defining the positive Hermitian operators ${\hat E}_{Dj}={\hat A}^{\dagger}_{j}{\hat 
A}_{j}$ and ${\hat E}_{F}={\hat A}^{\dagger}_{F}{\hat A}_{F}$.  It is also useful to define ${\hat E}_{D}=\sum_{j}{\hat A}^{\dagger}_{j}{\hat 
A}_{j}$.  The discrimination probability $P_{D}$ is constrained by the fact that ${\hat E}_{F}$ must be positive.  This, together with the decomposition of the identity, ${\hat E}_{F}+{\hat E}_{D}={\id}$, means that none of the eigenvalues of ${\hat E}_{D}$ may exceed unity.  It has been shown that the optimum measurement corresponds to the maximum eigenvalue of ${\hat E}_{D}$ being equal to 1\cite{Me1}.  

\section{Maximum discrimination probability for symmetric states}
\renewcommand{\theequation}{3.\arabic{equation}}
\setcounter{equation}{0}

In this section we derive the maximum attainable value of the unambiguous discrimination probability $P_{D}$ for symmetric states with equal {\em a priori} probabilities.  A set of quantum states $|{\psi}_{j}{\rangle}$ spanning a Hilbert space ${\cal H}$ is symmetric\cite{Ban} if there exists a unitary transformation ${\hat U}$ on ${\cal H}$ such that  
\begin{eqnarray}
|{\psi}_{j}{\rangle}&=&{\hat U}|{\psi}_{j-1}{\rangle}={\hat U}^{j}|{\psi}_{0}{\rangle}, \\ * |{\psi}_{0}{\rangle}&=&{\hat U}|{\psi}_{N-1}{\rangle}, \\ * {\hat U}^{N}&=&{\id}.
\end{eqnarray}
Eq. (3.3) follows from the fact that any state in ${\cal H}$ can be written as a superposition of the $|{\psi}_{j}{\rangle}$, and from ${\hat U}^{N}|{\psi}_{j}{\rangle}=|{\psi}_{j}{\rangle}$.  Such states are also said to be {\em covariant} with respect to ${\hat U}$ and have been found to have a preferential status with regard to problems in quantum detection theory.  In conventional quantum detection theory, where the aim is to maximise the discrimination probability or mutual information for all possible measurements and not just the subset defined by the no-errors constraint, the maximum discrimination probability for a set of symmetric states with equal {\em a priori} probabilities can be obtained exactly\cite{Helstrom,Ban}.  The optimum strategy uses the so-called `square root'\cite{Helstrom,Ban} or `pretty good'\cite{Hausladen} measurement.  Considerable progress has also been made towards maximising the mutual information for these states\cite{Osaki}, in particular in connection with symmetric quantum channels\cite{Sasaki}.  

As has been shown in \cite{Me1}, unambiguous discrimination between linearly-independent symmetric states arises naturally in connection with {\em entanglement concentration}, that is, transforming a fraction of an ensemble of systems all prepared in the same imperfectly entangled state into a maximally entangled state, using only local operations and classical communication.  The protocol given there, a generalisation of the 'Procrustean' technique due to Bennett {\em et al}\cite{Bennett}, will maximise the entanglement of a pair of subsystems with probability equal to the probability of discriminating between a certain set of linearly-independent symmetric states.  

Prior to solving for the maximum unambiguous discrimination probability for equally-probable linearly-independent symmetric states, we shall obtain a representation of them which simplifies our analysis.  We use the fact that the operator ${\hat U}$ can be expanded as follows:
\begin{equation}    
{\hat U}=\sum_{k=0}^{N-1}e^{i{\phi}_{k}}|{\gamma}_{k}{\rangle}{\langle}{\gamma}_{k}|,
\end{equation}
where ${\langle}{\gamma}_{k}|{\gamma}_{k'}{\rangle}={\delta}_{kk'}$.  The real angles ${\phi}_{k}$ may be taken to lie in half-open interval $[0,2{\pi})$.  It follows from Eq. (3.3) that
\begin{equation}
{\phi}_{k}=\frac{2{\pi}f_{k}}{N},
\end{equation}
where $f_{k}$ is an integer satisfying $0{\le}f_{k}{\le}n-1$.  It is convenient to arrange the $f_{k}$ in increasing order, so that $f_{k}{\ge}f_{k'}$ for $k>k'$.  Clearly, we can expand $|{\psi}_{0}{\rangle}$ as $\sum_{k}c_{k}|{\gamma}_{k}{\rangle}$, for some $c_{k}$ satisfying $\sum_{k}|c_{k}|^{2}=1$.  Together with Eqs. (3.1) and (3.4), this leads to
\begin{equation}
|{\psi}_{j}{\rangle}=\sum_{k=0}^{N-1}c_{k}e^{\frac{2{\pi}ijf_{k}}{N}}|{\gamma}_{k}{\rangle}.
\end{equation}
Note that the linear independence of the $|{\psi}_{j}{\rangle}$ implies that all of the $c_{k}$ are non-zero.  Linear independence means that no superposition of the $|{\psi}_{j}{\rangle}$ can vanish, so consider
\begin{equation}
\frac{1}{N}\sum_{j=0}^{N-1}e^{\frac{-2{\pi}ijr}{N}}|{\psi}_{j}{\rangle}=\sum_{k=0}^{N-1}c_{k}{\delta}_{rf_{k}}|{\gamma}_{k}{\rangle}.
\end{equation}
If the  $|{\psi}_{j}{\rangle}$ are linearly-independent, this must be non-zero for all $r=0,...N-1$.  Therefore, the $f_{k}$ must take every value in this range of integers.  As we have arranged these integers in increasing order, we find that $f_{k}$ is simply equal to $k$, so that linearly-independent symmetric states necessarily have the form:
\begin{equation}
|{\psi}_{j}{\rangle}=\sum_{k=0}^{N-1}c_{k}e^{\frac{2{\pi}ijk}{N}}|{\gamma}_{k}{\rangle}.
\end{equation}
That having this form is also a sufficient condition for linear independence is proven in \cite{Me1}.  For these states, the corresponding reciprocal states are given by
\begin{equation}
|{\psi}_{j}^{{\perp}}{\rangle}=Z^{-1/2}\sum_{r=0}^{N-1}c^{*-1}_{r}e^{\frac{2{\pi}ijr}{N}}|{\gamma}_{r}{\rangle} 
\end{equation}
where $Z=\sum_{r}|c_{r}|^{-2}$.  Note that the $|{\psi}_{j}^{{\perp}}{\rangle}$ are also symmetric, with respect to the same transformation ${\hat U}$ as the $|{\psi}_{j}{\rangle}$.

For symmetric states, the operator ${\hat E}_{D}$ has the explicit form
\begin{equation}
{\hat E}_{D}=\frac{1}{N^{2}}\sum_{j,r,r'}P_{j}c^{*-1}_{r'}c^{-1}_{r}e^{\frac{2{\pi}ij(r-r')}{N}}|{\gamma}_{r'}{\rangle}{\langle}{\gamma}_{r}|.
\end{equation}

Let us denote by ${\hat E}_{D}^{opt}$ an operator of this form which gives the maximum value of $P_{D}$.  We do not assume this operator to be unique, that is, we do not assume the optimum $P_{j}$ to be unique.  We can however, show that there exists an optimal operator ${\hat E}_{D}^{opt}$ which possesses the symmetry
\begin{equation}
{\hat E}_{D}^{opt}={\hat U}{\hat E}_{D}^{opt}{\hat U}^{\dagger}.
\end{equation}
We prove this by contradiction, by first supposing that no ${\hat E}_{D}^{opt}$ satisfies Eq. (3.11).  Consider now any operator ${\hat E}_{D}$ of the form (3.10).  If ${\hat E}_{D}$ corresponds to the maximum value of $P_{D}$, then its maximum eigenvalue, which we denote by ${\lambda}_{+}({\hat E}_{D})$, is equal to 1.  We then define
\begin{equation}
{\hat E}_{D}^{(l)}={\hat U}^{l}{\hat E}_{D}{\hat U}^{{\dagger}l}. 
\end{equation}
These operators clearly have the same eigenvalues as ${\hat E}_{D}$ and give the same value of $P_{D}$.  In fact, ${\hat E}^{(l)}_{D}$ can be obtained from ${\hat E}_{D}$ by cycling the probabilities $P_{j}$.  Writing explicitly the dependence of these operators on the $P_{j}$, we see that ${\hat E}^{(l)}_{D}(P_{j})={\hat E}_{D}(P_{j-l})$, where $P_{j{\pm}N}=P_{j}$.  Consider now the operator
\begin{equation}
{\hat E}_{D}^{ave}=\frac{1}{N}\sum_{l=0}^{N-1}{\hat E}_{D}^{(l)}=\frac{P_{D}}{N}\sum_{r}\frac{1}{|c_{r}|^{2}}|{\gamma}_{r}{\rangle}{\langle}{\gamma}_{r}|.
\end{equation}
The second equation here is true if the {\em a priori} probabilities ${\eta}_{j}$ are all equal to $1/N$, which we take to be the case.  The operator ${\hat E}_{D}^{ave}$ is invariant under the similarity transformation ${\hat E}^{ave}_{D}{\rightarrow}{\hat U}{\hat E}^{ave}_{D}{\hat U}^{\dagger}$ and gives the same transformation probability as ${\hat E}_{D}^{(l)}$.  Forming ${\hat E}^{ave}_{D}$ from ${\hat E}_{D}$ amounts to replacing all of the $P_{j}$ by their average value $P_{D}$, the quantity we wish to maximise.  However, its maximum eigenvalue ${\lambda}_{+}({\hat E}_{D}^{ave})$ satisfies 
\begin{equation}
{\lambda}_{+}({\hat E}_{D}^{ave}){\le}\frac{1}{N}\sum_{l}{\lambda}_{+}({\hat E}_{D}^{(l)})=1.
\end{equation}
This is a consequence of the fact the the maximum eigenvalue is convex on the space of Hermitian operators on ${\cal H}$.  A simple proof of this is given in the Appendix.  Let us define ${\hat E}_{D}^{'}={\hat E}_{D}^{ave}/{\lambda}_{+}({\hat E}_{D}^{ave})$.  The maximum eigenvalue of this operator is 1, so it is physically admissable.  However, the success probability for this operator is $P_{D}^{'}=P_{D}/{\lambda}_{+}({\hat E}_{D}^{ave}){\ge}P_{D}$, from (3.14).  Therefore, we can obtain from any  ${\hat E}_{D}$ another operator ${\hat E}_{D}^{'}$ which is invariant under the similarity transformation and whose associated discrimination probability $P^{'}_{D}$ is at least as high as $P_{D}$.  Thus, the premise that there is no ${\hat E}_{D}$ which gives the highest value of $P_{D}$ and has the specified symmetry is false.

It follows from Eq. (3.13), and from the orthogonality of the $|{\gamma}_{r}{\rangle}$, that the eigenvalues of ${\hat E}_{D}^{ave}$ are simply $P_{D}/N|c_{r}|^{2}$.  The optimum symmetric operator ${\hat E}^{opt}_{D}$ is simply that whose maximum eigenvalue is 1.  The desired least upper bound on $P_{D}$ is then given by
\begin{equation}
P_{D}{\le}N{\times}{\mathrm min}|c_{r}|^{2}.
\end{equation}
The bound here is clearly less than 1 unless all $|c_{r}|^{2}$ are equal to $N^{-1}$, in which case the $|{\psi}_{j}{\rangle}$ are orthogonal.  Although this inequality gives the analytic maximum discrimination probability, in practice it may be necessary to employ computational techniques to determine the smallest of the $|c_{r}|^{2}$, as we will see in the next section.  To find the maximum value of $P_{D}$ for a specific set of states, it is useful to have an expression for the $|c_{r}|^{2}$ which explicitly exhibits their dependence upon the states.  Using Eq. (3.8), we find that 
\begin{equation}
|c_{r}|^{2}=\frac{1}{N^{2}}\sum_{j,j'}e^{\frac{-2{\pi}ir(j-j')}{N}}{\langle}{\psi}_{j'}|{\psi}_{j}{\rangle}.
\end{equation}

The simplest set of linearly-independent symmetric states comprises just two states.  The problem of finding the maximum value of $P_{D}$ for a pair of states has been solved by Ivanovic\cite{Ivanovic}, Dieks\cite{Dieks} and Peres\cite{Peres1} for equal {\em a priori} probabilities and generalised by Jaeger and Shimony\cite{JShimony} to the case of unequal probabilities.  Denoting the two states by $|{\psi}_{\pm}{\rangle}$, the Ivanovic-Peres-Dieks limit for the probability of error-free state discrimination is
\begin{equation}
P_{IDP}=1-|{\langle}{\psi}_{+}|{\psi}_{-}{\rangle}|.
\end{equation}
It is interesting to see how this limit arises as a special case of the bound in (3.15).   Up to an irrelevant phase difference, the states $|{\psi}_{\pm}{\rangle}$ may be represented as 
\begin{equation}
|{\psi}_{\pm}>={\cos}{\theta}|+>{\pm}{\sin}{\theta}|->,
\end{equation}
where the angle ${\theta}$ lies in the range $[0,{\pi}/4]$ and the states $|{\pm}>$ constitute an orthogonal basis for the space spanned by $|{\psi}_{\pm}>$.  The system may be represented as a spin-1/2 
particle, and $|{\pm}>$ taken to be the eigenstates of ${\hat {\sigma}}_{z}$    
with eigenvalues ${\pm}1$.  Note that ${\hat {\sigma}}_{z}$ is the unitary operator ${\hat U}$ relating the states we aim to distinguish, since $|{\psi}_{\pm}>={\hat {\sigma}}_{z}|{\psi}_{\mp}>$.  We find that the corresponding reciprocal states are  
\begin{equation}
|{\psi}^{\perp}_{\pm}>={\sin}{\theta}|+>{\pm}{\cos}{\theta}|->.
\end{equation}
The expansion coefficents $c_{\pm}$ are given by $c_{+}={\cos}{\theta}$ and $c_{-}={\sin}{\theta}$.  Within the specified range of ${\theta}$, $|c_{+}|^{2}{\ge}|c_{-}|^{2}$, so that the maximum value of $P_{D}$ is $2|c_{-}|^{2}$, which is easily seen to be equal to $P_{IDP}$.

\section{Symmetric Coherent states}
\renewcommand{\theequation}{4.\arabic{equation}}
\setcounter{equation}{0}
In this section, we apply the bound (3.15) on $P_{D}$ to the problem of discriminating between symmetric coherent states of the harmonic oscillator or mode of a boson field.  These states are
\begin{equation}
|{\psi}_{j}{\rangle}=|{\alpha}_{j}{\rangle}=e^{ -\frac{|{\alpha}|^{2}}{2}}\sum_{n=0}^{\infty}\frac{{\alpha}_{j}^{n}}{\sqrt{n!}}|n{\rangle},
\end{equation}
where $j=0,...,N-1$ and ${\alpha}_{j}={\alpha}e^{\frac{2{\pi}ij}{N}}$, where ${\alpha}={\alpha}_{0}$ may be any complex number.  The $ |n{\rangle}$ are the usual boson number states.  The magnitudes of the complex arguments ${\alpha}_{j}$ are all equal to $|{\alpha}|$.  However their phases are distributed around the circle at regular intervals of $2{\pi}/N$.  Let us denote by ${\hat P}_{\cal H}$ the projector onto ${\cal H}$, the subspace spanned by the $|{\alpha}_{j}{\rangle}$.  The unitary transformation ${\hat U}$ which maps each state onto its successor is 
\begin{equation}
{\hat U}={\hat P}_{\cal H}e^{\frac{2{\pi}i{\hat n}}{N}}{\hat P}_{\cal H},
\end{equation}
where ${\hat n}$ is the boson number operator.  The quantities of interest if we wish to determine the maximum of $P_{D}$ are the square-moduli of the $c_{r}$.  One can show using Eq. (3.16) that
\begin{equation}
|c_{r}|^{2}=\frac{1}{N}\sum_{j}e^{\frac{-2{\pi}ijr}{N}}e^{|{\alpha}|^{2}(e^{\frac{2{\pi}ij}{N}}-1)}.
\end{equation}
Unfortunately, this summation seems to resist significant simplification, and in general must be carried out numerically.  A further complication arises if we wish to determine the maximum value of $P_{D}$, which entails finding the smallest of the $|c_{r}|^{2}$.  This is the fact that for general $N$, none of the $|c_{r}|^{2}$ remains the smallest for all values of $|{\alpha}|^{2}$.  This can be seen in figure 1, which shows the variation of the $|c_{r}|^{2}$ as functions of $|{\alpha}|^{2}$ for $N=10$.  The behaviour seen here for $N=10$ is typical of what happens for all $N$ except for $N=2$.  In this simplest case,   $|c_{0}|^{2}=(1+e^{-2|{\alpha}|^{2}})/2$ and $|c_{1}|^{2}=(1-e^{-2|{\alpha}|^{2}})/2$, so that $|c_{0}|^{2}{\ge}|c_{1}|^{2}$ for all $|{\alpha}|^{2}$.  For general $N$, each $|c_{r}|^{2}$ is less than the others for some range of $|{\alpha}|^{2}$.  At $|{\alpha}|^{2}=0$, we find that $|c_{0}|^{2}=1$ and all of the other $c_{r}$ are zero.  Our numerical results for various values of $N$ indicate that, as $|{\alpha}|^{2}$ increases, the smallest of the $|c_{r}|^{2}$ is successively $|c_{N-1}|^{2}$, then $|c_{N-2}|^{2}$ and so on until it is $|c_{0}|^{2}$ then the cycle repeats itself indefinitely.  It is evident from the figure that the point at which the minimum coefficient changes to a new one occurs when the derivative of the latter is zero.  This can be understood when we observe that the $|c_{r}|^{2}$ obey the relation  
\begin{equation}
\frac{d(|c_{r}|^{2})}{d(|{\alpha}|^{2})}=|c_{r-1}|^{2}-|c_{r}|^{2}.
\end{equation}
It follows that when the derivative of $|c_{r}|^{2}$ with respect to $|{\alpha}|^{2}$ is zero, we have $|c_{r}|^{2}=|c_{r-1}|^{2}$.  This is the point at which these functions cross and thus the smallest function ceases to be $|c_{r}|^{2}$ and becomes $|c_{r-1}|^{2}$.  As $|{\alpha}|^{2}{\rightarrow}{\infty}$, the $|c_{r}|^{2}$ tend to $1/N$.  Here, the overlaps between the states becomes indefinitely small, and the maximum discrimination probability approaches 1.

The maximum discrimination probability for $N=10$ is shown in figure 2.  We see that it is an increasing function of $|{\alpha}|^{2}$, although its derivative is discontinuous whenever a new $|c_{r}|^{2}$ becomes the smallest.

\section{Discussion}  

We have obtained the least upper bound on the unambiguous discrimination probability for linearly-independent symmetric states with equal {\em a priori} probabilities.  The corresponding detection operators ${\hat E}_{Dj}$ are  the simplest possible ones.  The only free parameters in these operators for general states are the conditional probabilities $P_{j}$.  For equally-probable symmetric states, the maximum value of $P_{D}$ is obtained when all $P_{j}$ are equal, and the one remaining free parameter is set by ${\lambda}_{+}({\hat E}_{D}^{opt})=1$, which is a necessary condition for the maximum value of $P_{D}$ for any linearly-independent set.  We thus find that for equally-probable symmetric states, the solution for the maximum of $P_{D}$ is the simplest $N$ state solution.  This appears also to be the case for optimisation problems in other areas of quantum detection theory, in particular the determination of the unconstrained maximum discrimination probability\cite{Helstrom,Ban}.  However, one issue we have not addressed here is whether or not it is only for equiprobable symmetric states that setting all $P_{j}$ to the same value gives the optimum measurement.

We have also examined the behaviour of the maximum $P_{S}$ for $N$ symmetric coherent states, in particular its dependence on the parameter $|{\alpha}|^{2}$, which has many interesting features for $N>2$.  Most notably, the derivative of the maximum discrimination probability is not continuous, owing to the fact that no single $|c_{r}|^{2}$ is smallest for all values of $|{\alpha}|^{2}$.  While this makes perfect sense from a mathematical point of view, the physical reasons for this pheneomenon are by no means obvious and further work may clarify the matter.

\section*{Acknowledgements}
We gratefully acknowledge financial support by the UK Engineering and Physical Sciences Research Council (EPSRC).

\section*{Appendix: Convexity of the maximum eigenvalue}
\renewcommand{\theequation}{A.\arabic{equation}}
\setcounter{equation}{0}

Consider a set of Hermitian operators ${\hat E}_{l}$ on a finite-dimensional Hilbert space ${\cal H}$, where $l=0,...,n-1$.  Let us denote by ${\lambda}_{+}({\hat E}_{l})$ the maximum eigenvalue of ${\hat E}_{l}$.  Then, as we show here, ${\lambda}_{+}$ is convex, that is, for any real, positive constants $a_{l}$,
\begin{equation}
{\lambda}_{+}(\sum_{l}a_{l}{\hat E}_{l}){\le}\sum_{l}a_{l}{\lambda}_{+}({\hat E}_{l}).
\end{equation}
This admits the following simple proof.  Let us write ${\hat S}_{r}=\sum_{l=0}^{r-1}a_{l}{\hat E}_{l}$ and let $|{s}_{r}{\rangle}$ be any eigenstate of ${\hat S}_{r}$ corresponding to the maximum eigenvalue ${\lambda}_{+}({\hat S}_{r})$.  Then,
\begin{eqnarray}
{\lambda}_{+}({\hat S}_{r+1})&=&{\langle}s_{r+1}|{\hat S}_{r}+a_{r}{\hat E}_{r}|s_{r+1}{\rangle} \nonumber \\ *
&{\le}&{\lambda}_{+}({\hat S}_{r})+a_{r}{\lambda}_{+}({\hat E}_{r}),
\end{eqnarray}
where the equality is satisfied only if ${\hat S}_{r}$ and ${\hat E}_{r}$ have a simultaneous eigenvector corresponding to the maximum eigenvalues of both of these operators.  From the definition of ${\hat S}_{r}$, we see that
\begin{equation}
a_{0}{\lambda}_{+}({\hat E}_{0})+\sum_{l=1}^{n-1}[{\lambda}_{+}(S_{l+1})-{\lambda}_{+}(S_{l})]={\lambda}_{+}(\sum_{i=0}^{n-1}a_{l}{\hat E}_{l}).
\end{equation}
Rearranging the inequality in (A.2) gives
\begin{equation}
{\lambda}_{+}({\hat S}_{l+1})-{\lambda}_{+}({\hat S}_{l}){\le}a_{l}{\lambda}_{+}({\hat E}_{l}).
\end{equation}
Substituting this inequality into (A.3) immediately gives (A.1), completing the proof.

\newpage

\begin{figure}

\epsfxsize=20cm

\epsfysize=20cm

\centerline{\epsffile{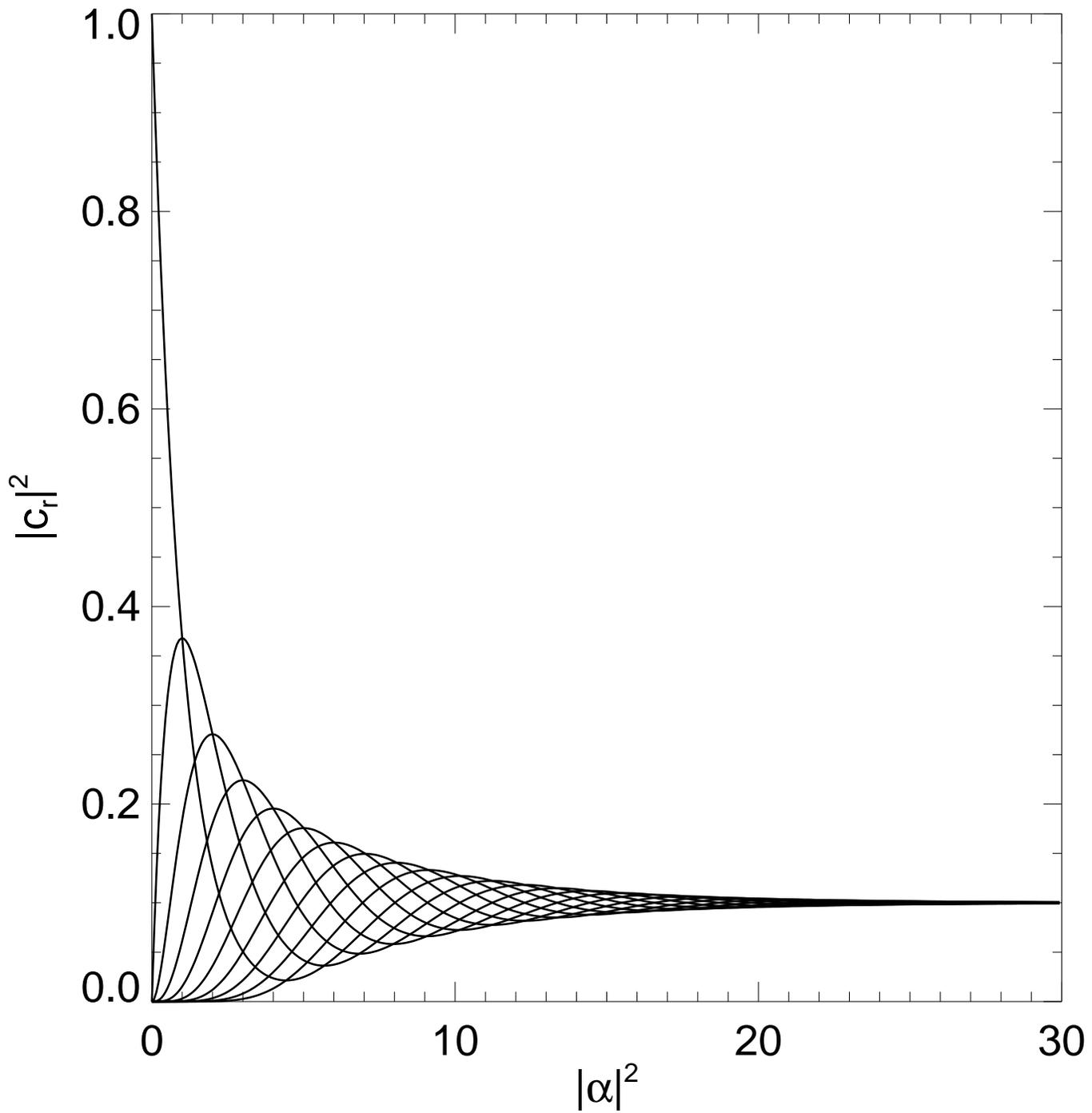}}

\caption{Dependence of the $|c_{r}|^{2}$ on $|{\alpha}|^{2}$ for 10 symmetric coherent states.}

\end{figure}

\begin{figure}

\epsfxsize=15cm

\epsfysize=15cm

\centerline{\epsffile{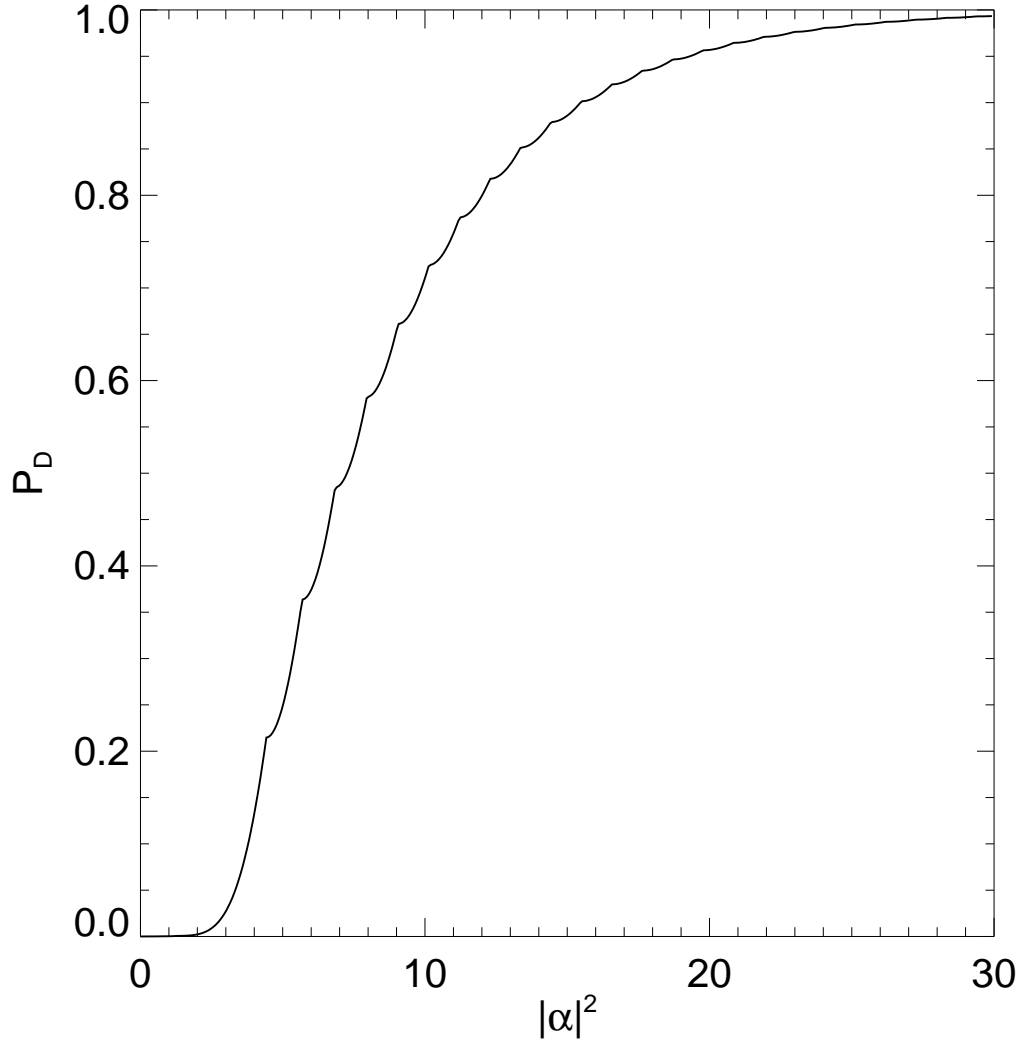}}

\caption{Maximum value of the probability $P_{D}$ of distinguishing between 10 symmetric coherent states as a function of $|{\alpha}|^{2}$.}

\end{figure}

\end{document}